\documentclass{article}
\usepackage{ijcai25}

\usepackage{times}
\usepackage{soul}
\usepackage{url}
\usepackage[hidelinks]{hyperref}
\usepackage[utf8]{inputenc}
\usepackage[small]{caption}
\usepackage{graphicx}
\usepackage{amsmath}
\usepackage{amsthm}
\usepackage{booktabs}
\usepackage{algorithm}
\usepackage{algorithmic}
\usepackage[switch]{lineno}

\usepackage{amsfonts}
\usepackage{bm}
\usepackage{amssymb}
\usepackage{wrapfig}
\usepackage{makecell}
\usepackage{color}
\usepackage{subcaption}

\usepackage{float}

\title{Constructive Conflict-Driven Multi-Agent Reinforcement Learning for Strategic Diversity}

\author{
Yuxiang Mai$^{1,2,3}$
\and
Qiyue Yin$^{1,2,3}$\and
Wancheng Ni$^{1,2,3,*}$\and
Pei Xu$^{2,3}$\and
Kaiqi Huang$^{1,2,3,}$\thanks{Corresponding authors}\\
\affiliations
$^1$School of Artificial Intelligence, University of Chinese Academy of Sciences\\
$^2$CRISE, Institute of Automation, Chinese Academy of Sciences\\
$^3$The Key Laboratory of Cognition and Decision Intelligence for Complex Systems, Institute of Automation, Chinese Academy of Sciences\\
\emails
\{maiyuxiang2020, wancheng.ni, pei.xu\}@ia.ac.cn,
\{qyyin, kqhuang\}@nlpr.ia.ac.cn
}
\begin{document}

\maketitle

\begin{abstract}
In recent years, diversity has emerged as a useful mechanism to enhance the efficiency of multi-agent reinforcement learning (MARL). However, existing methods predominantly focus on designing policies based on individual agent characteristics, often neglecting the interplay and mutual influence among agents during policy formation. To address this gap, we propose Competitive Diversity through Constructive Conflict (CoDiCon), a novel approach that incorporates competitive incentives into cooperative scenarios to encourage policy exchange and foster strategic diversity among agents. Drawing inspiration from sociological research, which highlights the benefits of moderate competition and constructive conflict in group decision-making, we design an intrinsic reward mechanism using ranking features to introduce competitive motivations. A centralized intrinsic reward module generates and distributes varying reward values to agents, ensuring an effective balance between competition and cooperation. By optimizing the parameterized centralized reward module to maximize environmental rewards, we reformulate the constrained bilevel optimization problem to align with the original task objectives. We evaluate our algorithm against state-of-the-art methods in the SMAC and GRF environments. Experimental results demonstrate that CoDiCon achieves superior performance, with competitive intrinsic rewards effectively promoting diverse and adaptive strategies among cooperative agents.\footnote{Codes are available at \href{https://github.com/YuxiangMai/CoDiCon}{https://github.com/YuxiangMai/CoDiCon}.}
\end{abstract}

\section{Introduction}
Due to the advancement of deep multi-agent reinforcement learning, many real-world problems have been modeled and addressed as cooperative multi-agent problems~\cite{samvelyan2019starcraft,sunehag2017value,rashid2020monotonic}, such as traffic signal control~\cite{wiering2000multi}, autonomous driving~\cite{hu2019interaction}, and robot control~\cite{DBLP:conf/rss/HaarnojaHZTTL19}. The goal of these cooperative multi-agent problems is to maximize the reward from a team perspective~\cite{colby2015counterfactual,rashid2020monotonic}. However, learning effective strategies for such complex multi-agent systems remains a significant challenge. One key problem is that relying on a single feedback signal often leads to homogeneous agent policies, resulting in inefficient exploration and hindering the emergence of complex cooperative behaviors~\cite{li2021celebrating}.\par

Providing additional reward signals~\cite{du2019liir,li2021celebrating,jiang2021emergence} to individual agents has proven to be an effective approach for diversifying cooperative strategies and improving performance. Existing methods for designing such rewards primarily rely on the mutual information between agent policies and agent identities (IDs) as intrinsic rewards~\cite{li2021celebrating,jiang2021emergence}, but such methods do not consider the influence of other agents and lack intuitive interpretability. LIIR~\cite{du2019liir} introduces learnable intrinsic rewards with shared parameters to facilitate information exchange, enabling agents to develop independent strategies. However, the process of learning intrinsic rewards overlooks the important role of competitive mechanisms in cooperative scenarios, as evidenced in sociological research.\par

There is a unique phenomenon observed in sociological research called Constructive Conflict~\cite{kirchmeyer1992multicultural,king2009conflict}. Constructive Conflict describes a situation in which the collective intelligence of a cooperative group is stimulated by the positive competition or clash of views between individuals and subgroups. Unlike destructive conflict that hinder the achievement of optimization goals, constructive conflict does not lead to strained relationships or reduced efficiency. Instead, it fosters diverse perspectives and strategy refinement, positively contributing to group goals. Inspired by the concept of constructive conflict, we propose competitive intrinsic rewards to enhance agent learning and team performance. In particular, unlike intrinsic rewards, the competitive intrinsic rewards provide cooperative agents with competitive incentives, thereby stimulating strategic communication and strategy diversification among agents.\par

In this paper, we propose an algorithm called CoDiCon, which is designed for competitive intrinsic rewards. Specifically, ranking is an effective method to encourage mutual competition, so we design an intrinsic reward that incorporates ranking property. The overall algorithm adopts an actor-critic structure~\cite{lowe2017multi,schulman2017proximal} based on MAPPO~\cite{yu2022surprising}, where each agent has a separate set of parameters. The global critic evaluates the current action values, and a centralized intrinsic reward generation module produces a ranked intrinsic reward for each agent's current action to foster competition. The agent's policy is jointly optimized using both extrinsic and intrinsic rewards, with the intrinsic reward serving merely as a training signal to distinguish rankings rather than having any inherent meaning. Mathematically, the optimization problem can be modeled as a constrained bilevel optimization problem, where the constraints in the outer optimization ensure the equivalence between the optimization objective and the original maximization of the environmental reward. We summarize
our contributions as follows:\par
\begin{itemize}
    \item We design an effective intrinsic reward mechanism based on the principles of constructive conflict, introducing competitive incentives to cooperative multi-agent systems to enhance team performance.
    \item The ranking module is proposed to provide agents with intrinsic rewards possessing ranking property. The optimization objective is formulated as a constrained bilevel problem, ensuring that optimizing the policy based on the ranking rewards aligns with the original goal of maximizing environmental rewards.
    \item Experimental results demonstrate that our algorithm outperforms existing methods. Visualizations of the agents' intrinsic rewards and state-reward space indicate that the learned intrinsic rewards produce distinct signals, enabling agents to take diverse actions collaboratively.
\end{itemize}

\section{Related Work}
In recent years, deep multi-agent reinforcement learning has made significant advancements. Research efforts such as COMA~\cite{foerster2018counterfactual}, MADDPG~\cite{lowe2017multi}, and LICA~\cite{zhou2020learning} have explored policy-based approaches to multi-agent problems, utilizing a centralized critic to evaluate the value of each distributed policy. Value decomposition methods, including VDN~\cite{sunehag2017value}, QMIX~\cite{rashid2020monotonic}, and QTRAN~\cite{hostallero2019learning}, decompose environmental feedback into individual agent value functions, thereby enabling credit assignment. QPLEX~\cite{wang2020qplex} proposed using dueling networks to relax the monotonicity constraint of mixing networks, expanding their representational capacity. Learning from others is an innate human survival skill, a phenomenon mirrored in agent policies, where information exchange via mixing networks enhances policy generation. However, most existing methods focus on centralized mixing network structures to satisfy value function constraints, often relying on strong assumptions about these functions. In contrast, our approach learns explicit intrinsic rewards for each agent at each time step, avoiding such assumptions and enabling immediate credit assignment.\par

Our work addresses the problem of designing intrinsic rewards in cooperative multi-agent systems, a topic that has been explored in some previous studies. EOI~\cite{jiang2021emergence} proposed learning a classifier for observations to compute the probability that an observation belongs to each agent, using this probability as an intrinsic reward to adjust the agent's final reward. CDS~\cite{li2021celebrating} introduced maximizing the mutual information between agent IDs and trajectories as an intrinsic reward to promote diverse policies. However, such methods do not consider the influence of other agents, overlooking the importance of learning from them, and they lack intuitive interpretability. LIIR~\cite{du2019liir} designed a module for generating intrinsic rewards to guide more diverse policies, ensuring that the two-level optimization is equivalent to the initial optimization problem. However, it also overlooks the positive competitive influence that other agents can have on policy development. Furthermore, directly calculating intrinsic rewards can result in an excessively large optimization space, increasing the risk of policies converging to local optima.

\section{Preliminary}\label{sec:pre}
\subsection{Cooperative Multi-Agent Reinforcement Learning}
Multi-agent reinforcement learning~\cite{sutton2018reinforcement} extends traditional reinforcement learning to address decision-making problems with multiple agents in sequential environments. A fully cooperative multi-agent problem can be represented as a Decentralized Partially Observable Markov Decision Process (Dec-POMDP)~\cite{oliehoek2016concise}. In this model, a Dec-POMDP is described by the tuple $\langle S, A, U, Z, O, P, r, n, \gamma \rangle$. Here, $S$ represents the global states of the environment, and $A$ denotes the set of $n$ agents. At each time step $t$, each agent $i \in A$ selects an action $u_i$ from its action set $U_i$, resulting in a joint action $\bm{u} \in \bm{U} \equiv U^n$. The state transition function $P$ determines the next state $s' \in S$ based on the current state $s$ and the joint action $\bm{u}$. All agents share a common reward $r(s, \bm{u})$, which depends on the state and joint action. A discount factor $\gamma \in [0, 1)$ is used to weigh future rewards. In the partially observable setting, each agent receives a local observation $z_i \in Z$, derived from the observation function $O(s, \bm{u}): S \times \bm{U} \rightarrow Z$, where $Z$ is the observation space. Each agent learns an individual policy $\pi_i(u_i | \tau_i; \theta_i)$ with parameters $\theta_i$ based on its own action-observation history $\tau_i$. The agents' combined policies determine the joint action taken in the environment.

\begin{figure*}
    \centering
    \includegraphics[width=0.7\linewidth]{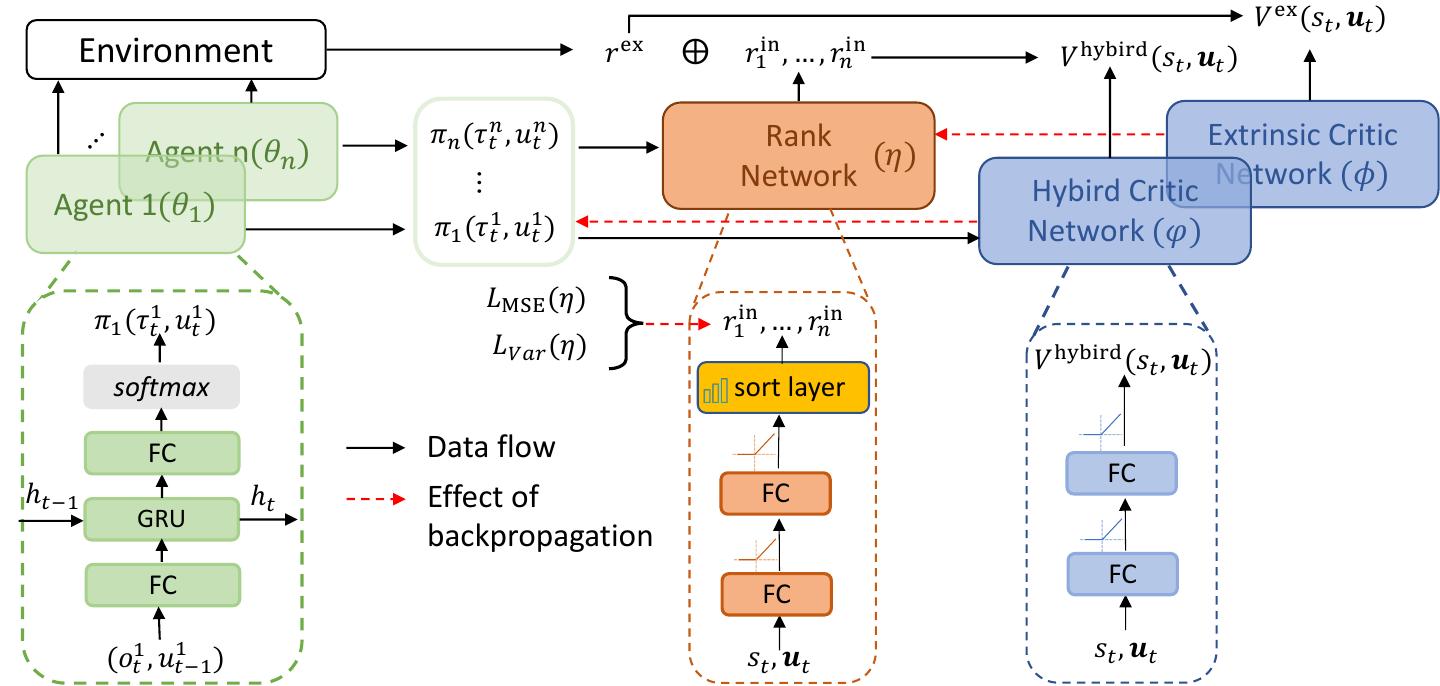}
    \caption{Framework of the proposed method. The framework consists of four parameterized modules: the agent module with parameters $\bm{\theta}$, the ranking module with parameters $\eta$, the hybird critic module with parameters $\varphi$, and the extrinsic critic module with parameters $\phi$.}
    \label{fig:overview}
\end{figure*}

\subsection{Centralized Training with Decentralized Execution}
Centralized Training and Decentralized Execution (CTDE) is a commonly used paradigm for addressing multi-agent problems. In this paradigm, actor-critic methods are often chosen~\cite{williams1992simple,andrychowicz2016learning,schulman2015trust,sutton2018reinforcement,yu2022surprising}. In our approach, we choose the MAPPO~\cite{yu2022surprising} algorithm as the base framework, which consists of $n$ independently parameterized policies $\pi_{\theta_i}$. The policy parameters $\bm{\theta}=\{\theta_1,\theta_2,...,\theta_n\}$ are optimized using policy gradient methods to maximize the extrinsic reward:
\begin{align}
    J^{\text{ex}}(\bm{\theta})=\mathbb{E}_{s, \bm{u}} \left[ \min \left(p(\bm{\theta})A_{\pi}\left(s, \bm{u}\right),\widetilde{p}(\bm{\theta})A_{\pi}\left(s, \bm{u}\right) \right)\right] \label{eq:partial},
\end{align}
where $p(\bm{\theta})=\frac{\pi_{\bm{\theta}}(u|\bm{s})}{\pi_{\text{old},\bm{\theta}}(u|\bm{s})}$ denotes the policy ratio, which depends on the policy parameters $\bm{\theta}$, and $\widetilde{p}(\bm{\theta})=\text{clip}(p(\bm{\theta}),1-\epsilon,1+\epsilon)$ is the clipped policy ratio. $A_{\pi}(s, \bm{u})$ represents the advantage function based on the state and actions. There are several methods to estimate $A_\pi(s,\bm{u})$. For example, $A_{\pi}(s, \bm{u}) = r^{\text{ex}}(s, \bm{u}) + V^{\text{ex}}(s^\prime) - V^{\text{ex}}(s)$ is the standard advantage function~\cite{schulman2015trust}, where $s^\prime$ is the next state.

\section{Method}
In this section, we present the competitive intrinsic reward algorithm, CoDiCon. First, we provide a formal definition of the problem and represent it as a constrained bilevel optimization problem. Then, we detail the approach for solving this problem, which effectively addresses the interplay between agents' learning and their competitive interactions.
\subsection{The Optimization Objective}
We define the components of the agent's reward, which consist of intrinsic reward $r^{\rm in}$ and extrinsic reward $r^{\rm ex}$. The intrinsic reward is parameterized by $\eta$. At each time step, each agent takes a state-action pair as input, the hybird reward can be expressed as:
\begin{align}
    r_{i,t}^{\text{hybird}} = r_{t}^{\text{ex}} + \lambda r_{i,t}^{\text{in}}(\eta) \label{r_hybird}.
\end{align}
In equation~(\ref{r_hybird}), $\lambda$ represents the hyperparameter that balances intrinsic and extrinsic rewards. It is important to note that the additional intrinsic reward does not appear in the standard multi-agent problem. After defining a hybird reward $r_{i,t}^{\text{hybird}}$ for each agent at each time step, we define the discounted hybird reward for each agent as follows:
\begin{align}
    R_{i,t}^{\text{hybird}} = \sum_{l=0}^{\infty} \gamma^l \left( r_{t+l}^{\text{ex}} + \lambda r_{i,t+l}^{\text{in}}(\eta) \right),
\end{align}
and the hybird value function for agent $i$ is defined as:
\begin{align}
    V_{i}^{\text{hybird}}(s_{t}) = \mathbb{E}_{s_{t+1},u_{i,t}, \ldots} \left[ R_{i,t}^{\text{hybird}} \right].
\end{align}
Unlike the extrinsic value function $V^{\text{ex}}$, these hybird value functions $V_i^{\text{hybird}}$ do not have any actual physical meaning. They are just used to update the parameters $\theta_i$ of each agent's policy. Next, we consider the overall optimization objective, defined as:
\begin{align}
\max_{\eta, \bm{\theta}} \quad & J^{\text{ex}}(\eta) , \label{eq:proj} \\
\text{s.t.} \quad & \theta_i = \arg\max_{\bm{\theta}} J^{\text{hybird}}_i(\bm{\theta}, \eta), \quad \forall i \in [1, 2, \ldots, n],  \nonumber \\
& r_1^{\text{in}} < r_2^{\text{in}} < \cdots < r_n^{\text{in}}, \quad \text{where} \quad r_i^{\text{in}} = \text{rank}(\eta) \nonumber
\end{align}
where $J^{\text{hybird}}_i=\mathbb{E}_{s_{0},u_{i,0},...}\left[R_{i,0}^\text{hybird}\right]$ depending on $\theta_i$ and $\eta$, $\eta$ indicates the parameter of the ranking module and $\bm{\theta}$ indicates the policy parameter set $\{\theta_1, \theta_2,\cdots,\theta_n\}$. The updates of the ranking module parameters and policy parameters are performed alternately. The ranking network module consists of fully connected layers and a sorting function layer, ensuring that the intrinsic rewards $r^\text{in}$ it outputs have numerical differences and are arranged in ascending order. When the parameters of the ranking module are frozen, the policy parameter $\theta_i$ is optimized by maximizing the hybrid expected cumulative return $J_i^\text{hybird}$ for agent $i$. The advantage of this approach lies in using the ranked intrinsic rewards at each step for policy learning, where competition is captured by the intrinsic rewards, defined through pairwise performance ranking as $r_i^\text{in}=\text{rank}_\eta(i, -i)$. This mechanism motivates agents to compete and learn from each other, leading to more complex and diverse behaviors. Conflict arises when an agent's action that maximizes its intrinsic reward $r_i^\text{in}$ opposes the team’s extrinsic reward $r^\text{ex}$. Specifically, conflict exists if for some agent $i$, $\nabla_{u_i} r^\text{ex} \cdot \nabla_{u_i} r_i^\text{in} < 0$, implying the agent’s optimal behavior hinders team performance. From an optimization perspective, problem~(\ref{eq:proj}) can be viewed as a constrained bilevel optimization problem, where the outer optimization constrains the inner policy improvement process. The ultimate goal of optimization is to maximize the environmental reward under the constraint of intrinsic rewards. This ensures alignment with the original objective of maximizing the environmental reward, allowing better-performing policies to receive higher intrinsic rewards. In the next section, we will discuss the relationship between $J^{\text{ex}}$ and the intrinsic reward parameter $\eta$ during the optimization process.
\subsection{Algorithm}
For the constrained two-level optimization problem, at each update step, the policy parameters are updated based on the hybrid expected return, while the update of the ranking module parameters depends on the ranking process and the extrinsic expected return.

Specifically, the parameters of each agent are updated using the hybird critic network. Given the trajectory data generated by the agents, the policy parameters are updated through the policy gradient method as described in~(\ref{eq:partial}):
\begin{align}
    \nabla_{\theta_i} \min \left(p(\theta_i)A_i^\text{hybird}\left(s, \bm{u}\right),\widetilde{p}(\theta_i)A_i^\text{hybird}\left(s, \bm{u}\right) \right), \label{eq:a_hybird}
\end{align}
where $p(\theta_i)$ is the policy ratio of agent $i$. $A_i^{\text{hybird}}(s,\bm{u})$ denotes the hybird critic, which can be chosen in various ways~\cite{sutton2018reinforcement,schulman2015trust,yu2022surprising}. In this paper, we choose $A_i^{\text{hybird}}(s,\bm{u})=r_i^\text{hybird}(s,\bm{u})+V_\varphi^\text{hybird}(s^\prime)-V_\varphi^\text{hybird}(s)$ as the advantage function, where $V_\varphi^\text{hybird}$ represents the hybird state value, parameterized by $\varphi$, and $s^\prime$ denotes the next state of the agent in the trajectory. Given~(\ref{eq:a_hybird}) and a policy learning rate $a$, the updated policy parameter $\theta_i^\prime$ can be expressed as: $\theta_i^\prime=\theta_i+\alpha\nabla_{\theta_i} \min \left(p(\theta_i)A_i^\text{hybird}\left(s, \bm{u}\right),\widetilde{p}(\theta_i)A_i^\text{hybird}\left(s, \bm{u}\right) \right)$. 

To ensure that intrinsic rewards promote competition among agents, we update $\eta$ by applying a relative numerical constraint to the intrinsic rewards produced by the ranking module. This ensures that the rewards follow a sequential distribution. Specifically, we use mean squared error (MSE) loss and variance loss of intrinsic rewards to maintain the output as a sequence with numerical differences:
\begin{align}
    \mathcal{L}_\text{MSE}(\eta) = \frac{1}{n} \sum_{i=1}^{n} (r^\text{in}_i - y_i)^2, \label{eq:mse}
\end{align}
\begin{align}
    \mathcal{L}_\text{Var}(\eta) = \frac{1}{n} \sum_{i=1}^{n} (r^\text{in}_i - \bar{r}^{\text{in}})^2,
\end{align}
where $y$ is the optimization target for the intrinsic reward, represented as an ordered sequence. This sequence is randomly initialized at the beginning of training (in our setup, 20\% positive values and 80\% negative values) and remains fixed during subsequent training. $\bar{r}^{\text{in}}$ is the mean of the intrinsic rewards. For the update of the parameter $\eta$, we first minimize the MSE loss and maximize the variance loss:
\begin{align}
    \mathcal{L}(\eta)=\beta_1\mathcal{L}_\text{MSE}(\eta)-\beta_2\mathcal{L}_\text{Var}(\eta). \label{eq:rank_loss}
\end{align}
Given~(\ref{eq:rank_loss}) and a learning rate $\beta$, the updated parameters of the ranking module can be expressed as: $\eta' = \eta - \beta \nabla_{\eta} \mathcal{L}(\eta)$.

Next, we construct expressions for $\eta'$ and $J^\text{ex}$ to update the parameter $\eta'$. Using the updated policy parameter $\bm{\theta}^\prime$, we apply the chain update rule to get:
\begin{align}
    \nabla_{\eta^\prime} J^{\text{ex}} = \nabla_{\theta_i^\prime} J^{\text{ex}} \nabla_{\eta'} \theta_i'. \label{eq:j_ex}
\end{align}
The purpose of~(\ref{eq:j_ex}) is to formally express the impact of $J^\text{ex}$ on the updated policy parameter $\eta^\prime$ through the updated parameter $\theta_i^\prime$. This technique is widely adopted in meta-gradient learning~\cite{andrychowicz2016learning,santoro2016meta,xu2018meta}. Using samples generated by the updated policy network, the meta-gradient $\nabla_{\eta^\prime}$ can be computed. In~(\ref{eq:j_ex}), $\nabla_{\theta_i^\prime} J^{\text{ex}}$ can be estimated by stochastic gradient as
\begin{align}
    \nabla_{\theta_i'} \min \left(p(\theta_i')A^\text{ex}\left(s, \bm{u}\right),\widetilde{p}(\theta_i')A^\text{ex}\left(s, \bm{u} \right)\right), \label{eq:adv}
\end{align}
here, $A^\text{ex}(s,\bm{u})$ denotes the centralized extrinsic critic. Similar to the centralized hybrid critic, we define $A^\text{ex}(s,\bm{u})=r^\text{ex}(s,\bm{u})+V_\phi^\text{ex}(s^\prime)-V_\phi^\text{ex}(s)$, where $V_\phi^\text{ex}(s)$ is the extrinsic value function with parameters $\phi$. For the update of $\eta^\prime$, the second term in~(\ref{eq:j_ex}) can be derived as:
\begin{align}
    \nabla_{\eta'} \theta_i' = &\nabla_{\eta'} \big{[}\theta_i + \alpha \nabla_{\theta_i} \min\left(p(\theta_i)A_i^{\text{hybird}},\widetilde{p}(\theta_i)A_i^\text{hybird}\right) \big{]} \nonumber\\
    = &\alpha \lambda \min(\nabla_{\theta_i}p(\theta_i)\nabla_{\eta'}r_i^{\text{hybird}},\nabla_{\theta_i}\widetilde{p}(\theta_i)\nabla_{\eta'}r_i^{\text{hybird}}). \label{eq:second_term}
\end{align}
Figure~\ref{fig:overview} presents the overall framework of the CoDiCon algorithm. A detailed description of the algorithm is provided in Algorithm~\ref{alg:algorithm}.
\begin{algorithm}[t]
    \caption{The algorithm of CoDiCon.}
    \label{alg:algorithm}
    \begin{algorithmic}[1] %[1] enables line numbers
    \STATE \textbf{Input:} Policy learning rate $\alpha$ and intrinsic reward learning rate $\beta$.
    \STATE \textbf{Initialize:} Policy parameters $\bm{\theta}$ and intrinsic reward parameters $\eta$.
    \FOR{$t=1$ to $T_{\text{max}}$}
    \STATE Sample a trajectory $\mathcal{D}=\{s_0,\bm{u_0},s_1,\bm{u_1},\cdots\}$ by executing actions with the decentralized policies $\{\pi_{\theta_1},\cdots,\pi_{\theta_n}\}$;
    \STATE Update $\bm{\theta}$ according to~(\ref{eq:a_hybird}) with learning rate $\alpha$;
    \STATE Update $\eta$ according to~(\ref{eq:rank_loss}) with learning rate $\beta$;
    \STATE Compute (\ref{eq:adv}) and (\ref{eq:second_term}) using samples from $\mathcal{D}$ after the first update of $\eta$;
    \STATE Update $\eta^\prime$ according to~(\ref{eq:j_ex}) and step 7 with learning rate $\beta$;
    \ENDFOR
    \STATE \textbf{return} policy parameters $\bm{\theta}$
    \end{algorithmic}
    \end{algorithm}

\section{Experiments}
\begin{figure*}
    \centering
    % 第一张图片
    \begin{subfigure}[b]{0.22\textwidth} % 设置宽度
        \centering
        \includegraphics[width=\textwidth]{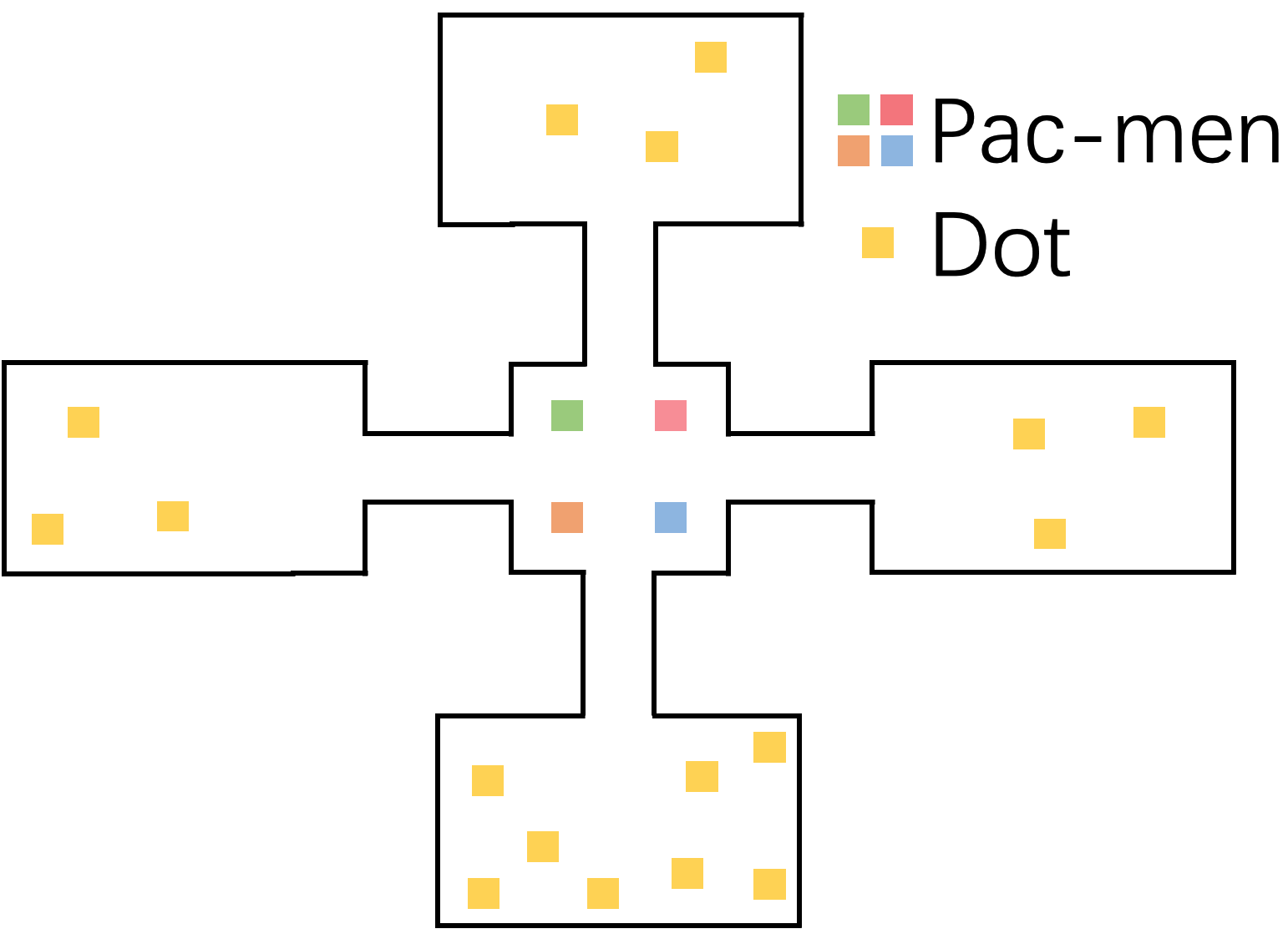} % 图片路径
        \caption{}
        \label{fig:pacmen1}
    \end{subfigure}
    % 第二张图片
    \begin{subfigure}[b]{0.22\textwidth}
        \centering
        \includegraphics[width=\textwidth]{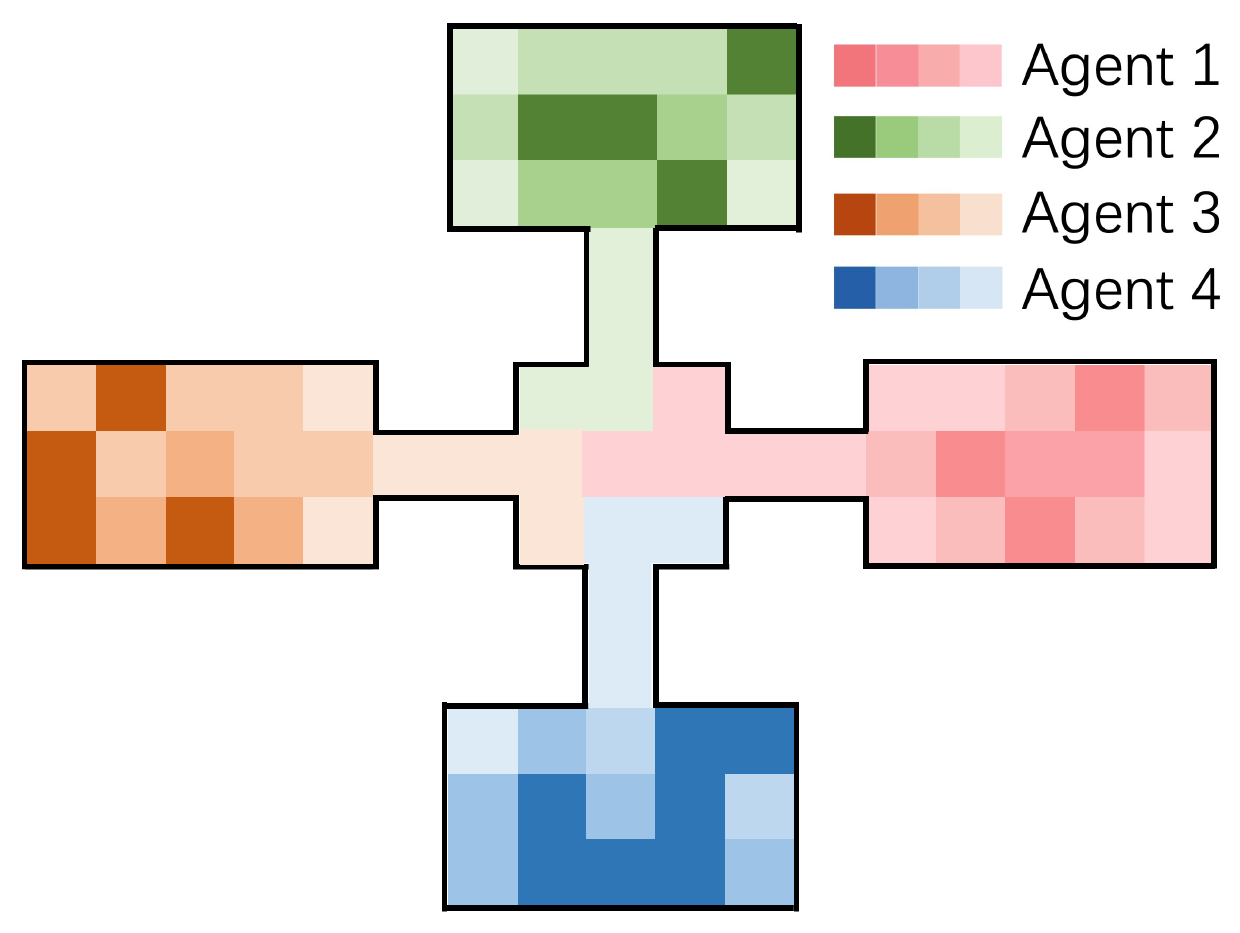} % 图片路径
        \caption{}
        \label{fig:pacmen2}
    \end{subfigure}
    % 第三张图片
    \begin{subfigure}[b]{0.22\textwidth}
        \centering
        \includegraphics[width=\textwidth]{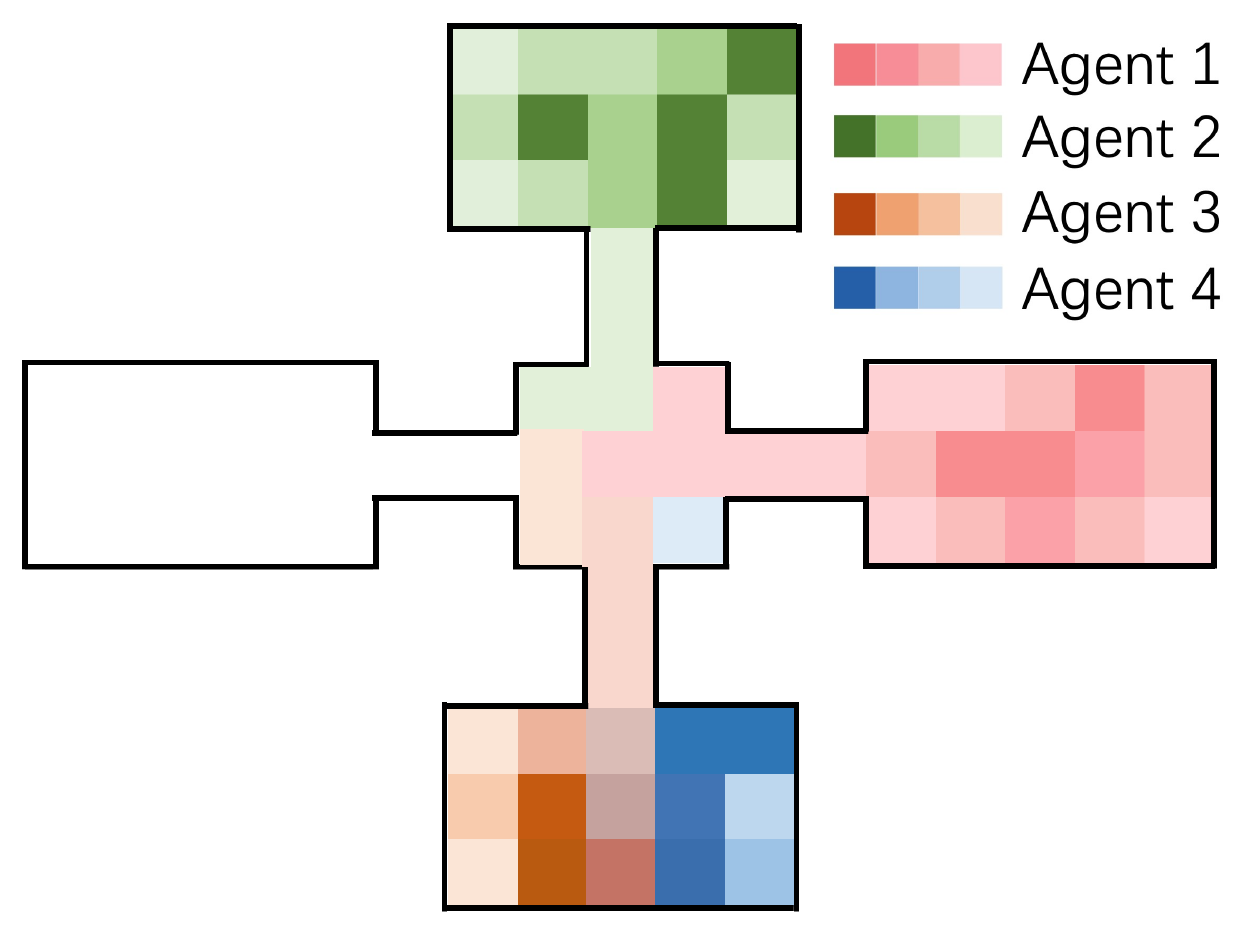} % 图片路径
        \caption{}
        \label{fig:pacmen3}
    \end{subfigure} 
    % 第四张图片
    \begin{subfigure}[b]{0.21\textwidth}
        \centering
        \includegraphics[width=\textwidth]{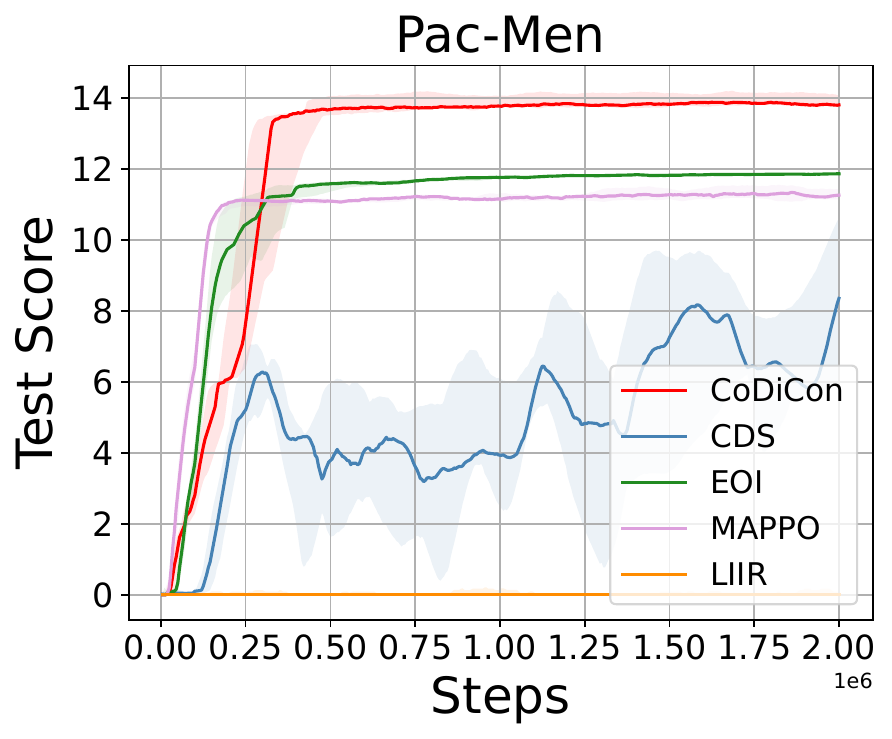} % 图片路径
        \caption{}
        \label{fig:pacmen4}
    \end{subfigure}    
    \caption{(a) The designed Pac-Men environment. (b) Visitation heatmap of EOI. (c) Visitation heatmap of our CoDiCon. The darker color means the higher value. (d) The training curve for the compared methods.}
    \label{fig:competitive_pacmen}
\end{figure*}

\begin{figure*}
    \centering
    \includegraphics[width=0.72\linewidth]{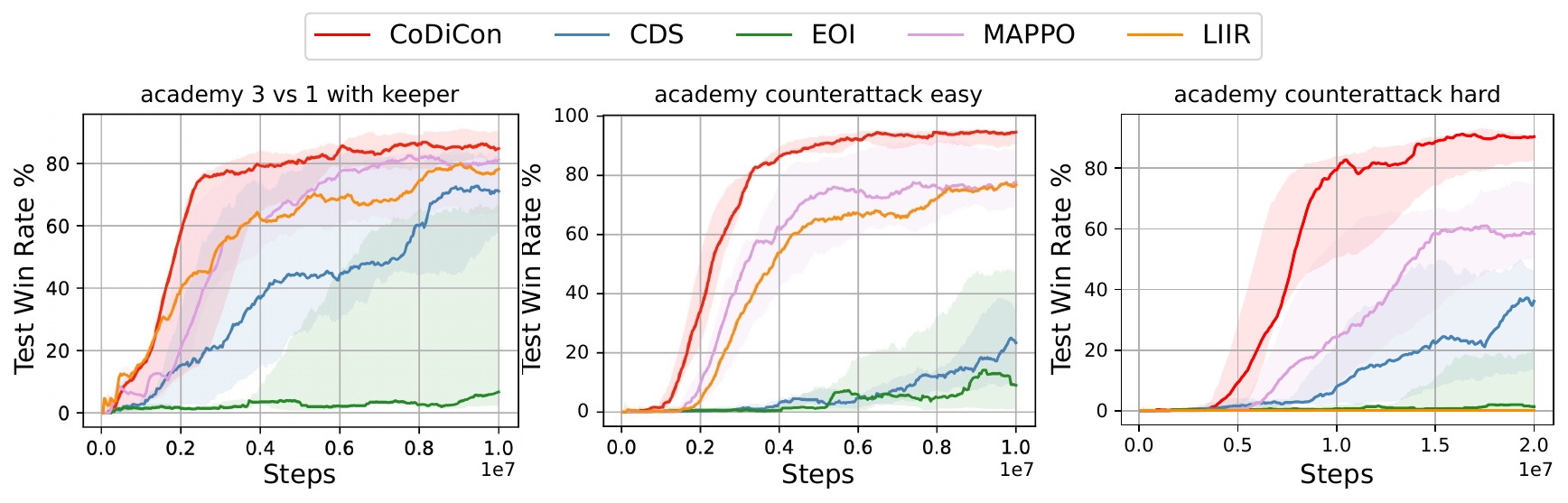}
    \caption{Training curves compared with the baselines on GRF.}
    \label{fig:results_grf}
\end{figure*}

In this section, we analyze and illustrate the performance and effectiveness of our algorithms in Pac-Men, Google Reasearch Football (GRF) and Starcraft Multi-agent Changellenge (SMAC) environments. We select the multi-agent policy-based method (MAPPO), the intrinsic reward policy-based method (LIIR), and the independent emergence methods (EOI, CDS) as baseline methods for comparison. First, we analyze the effectiveness of introducing a competitive mechanism to improve policy performance in the Pac-Man environment. Next, we conduct performance comparison experiments on GRF and SMAC. Then, we perform a case study on GRF to analyze the effectiveness of the intrinsic rewards in our algorithm. Additionally, we use t-SNE visualization in the SMAC environment to compare the final distribution of policies in the state-reward space, explaining the intrinsic reasons behind the effectiveness of the algorithm. Finally, we experimentally validate the two modules of the algorithm that encourage intrinsic reward variability through ablation studies.

\subsection{Competitive Intrinsic Reward}
We design the Pac-Men environment, shown in Figure~\ref{fig:pacmen1}, to demonstrate how our algorithm encourages agents to compete, thereby improving overall algorithm performance. In this environment, the central room connects four equally sized rooms and is equidistant from each of them. Four agents are initialized at fixed positions near the entrance of each room, and each agent has a 5x5 field of view. To foster competition among agents, the lower room contains the highest number of "dots" compared to the other three rooms. To make the environment more challenging, the game is limited to 17 timesteps, making it difficult for a single agent to collect all dots in the lower room, requiring at least two agents. The best case scenario is that one of the agents gives up the closest room and chooses to enter the room below, and then the two agents eat all the dots in the room below together. Agents incur a -0.25 penalty per timestep but gain 1 reward for each bean eaten. We compare our algorithm with SOTA algorithms. As shown in the visitation heatmap in Figure~\ref{fig:pacmen2} and~\ref{fig:pacmen3}, the EOI agents dispersed to four different rooms, whereas two agents in our algorithm both moved to the lower room, which contains the highest number of dots. In this environment, overly dispersed strategies do not result in optimal rewards. The experimental results shown in Figure~\ref{fig:competitive_pacmen} demonstrate that our algorithm successfully discovers the optimal strategy, while other algorithms fall into local optimal solutions. The possible reason is that our algorithm enables other agents to learn the high-reward strategy of the agent that moves to the lower room to collect "dots" by continuously ranking the intrinsic rewards, that is, by assigning higher intrinsic rewards to the agents that go to the lower room.

\subsection{Performence on GRF}
In this section, we first evaluate the performance of the algorithm in the GRF environment. Specifically, we compare our algorithm with others on GRF tasks of increasing difficulty, including  academy\_3\_vs\_1\_with\_keeper, academy-\_counterattack\_easy, and academy\_counterattack\_hard. In the GRF task, the agents need to collaborate in time and space to organize the attacking opportunities, and only the scoring are rewarded. In our experiments, we control the agents on the left (in yellow) except the goalkeeper's agent, and the agents on the right are controlled via the roule-based AI built into the game engine. The agents have 19 discrete action spaces, including moving from different angles, sliding, shooting, and passing. Observations include the position and direction of movement of the agents and other agents, as well as the position and direction of movement of the ball. The movement of the soccer ball in the z-axis direction is also included in the observations, and the agents receive feedback from the environment only at the end of the game.

\begin{figure*}
    \centering
    \includegraphics[width=0.7\linewidth]{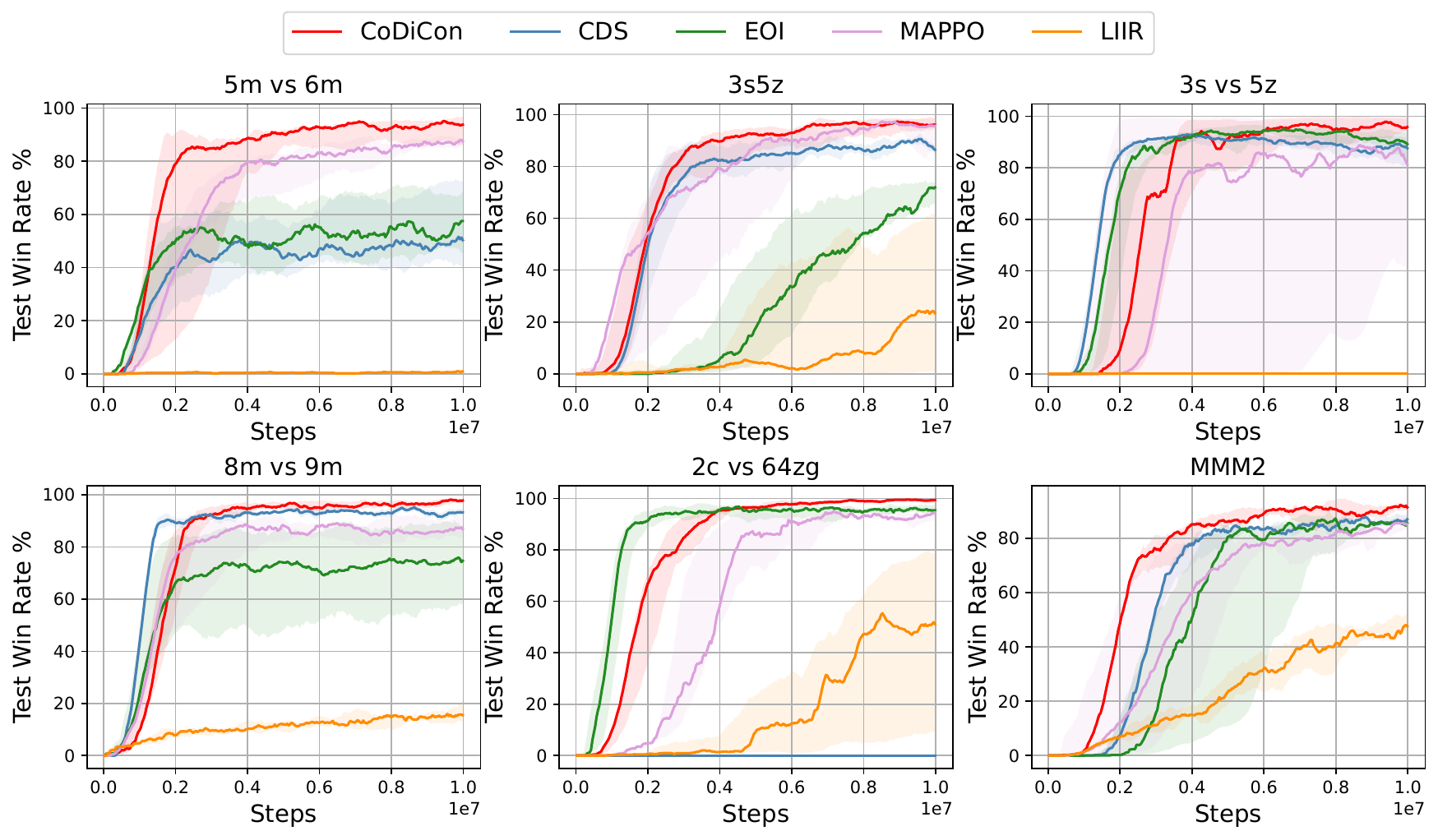}
    \caption{Training curves compared with the baselines on SMAC.}
    \label{fig:results_smac}
\end{figure*}

Our algorithm outperforms all other algorithms in all test environments, and the results are shown in Figure~\ref{fig:results_grf}. In the academy\_3\_vs\_1\_with\_keeper, other baselines also perform well, such as MAPPO and LIIR, showing performance close to that of our algorithm. However, on the progressively more difficult environments of academy\_counterattack\_easy and academy\_counterattack\_hard, the advantage of our algorithm gradually widens, and the interval between the training curves becomes more and more obvious. This demonstrates that our algorithm is able to effectively capture valuable states and actions in sparse reward-difficult environments, generating intrinsic rewards that encourage the agent to learn better.
\subsection{Performence on SMAC}
In this section, we test our algorithm in the StarCraft Micromanagement (SMAC) environment, a popular MARL benchmark. Each agent has attributes like health, weapon CD, shield, unit type, last action, and relative distances to observed units, with enemy units sharing similar attributes except CD. In partially observable settings, agents gather information within a circular range. The action space includes four movement directions, $k$ attack actions (where $k$ is the max number of enemies), a stop action, and a no-op action, with invalid actions masked. We evaluate on scenarios: 3s\_vs\_5z, 8m\_vs\_9m, MMM2, 2c\_vs\_64zg, 5m\_vs\_6m, 3S5Z. Only the Zealot is melee, the Medivac is a non-attacking support unit, and others are ranged. Agents receive a joint team reward based on total damage, with a significant bonus for winning.

We present the experimental results of the comparison in Figure~\ref{fig:results_smac}. Our algorithm outperforms the baseline algorithms in most environments, with performance comparable to EOI only in the environment of 2c\_vs\_64zg. Both CDS and EOI achieve performance close to our algorithm on some maps, such as 3s\_vs\_5z and 8m\_vs\_9m. However, their performance on 5m\_vs\_6m and 2c\_vs\_64zg are highly unstable, indicating that the diversity mechanisms in CDS and EOI lack effective exploration and utilization of the state space, which may lead to local optima. MAPPO shows relatively stable and high performance across all environments, but it still lags behind our algorithm in terms of training efficiency and convergence speed. In contrast, LIIR performs poorly on all SMAC maps, suggesting that the unconstrained learnable intrinsic rewards in LIIR make learning highly unstable, particularly in scenarios with denser rewards. In contrast, our algorithm provides stable initial feedback during training by introducing a distinctly different prior intrinsic reward.

\subsection{Visualizing the Learned Intrinsic Reward}
\begin{figure}
    \centering
    % 第一张图片
    \begin{minipage}[c]{0.25\textwidth} % 调整宽度
        \centering
        \includegraphics[width=\textwidth]{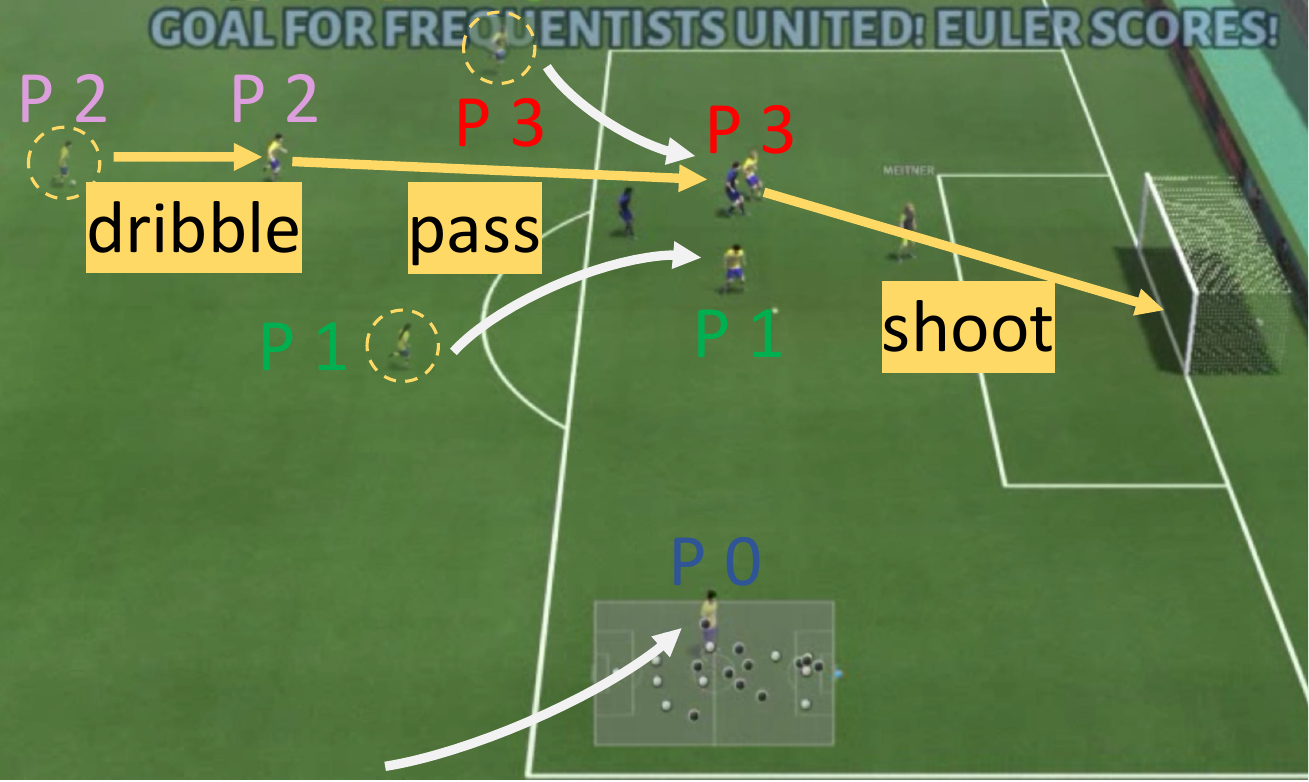}% 图片路径
        % \label{fig:grf_case}
    \end{minipage}
    \hfill % 设置两张图片之间的距离
    % 第二张图片
    \begin{minipage}[c]{0.21\textwidth} % 调整宽度
        \centering
        \includegraphics[width=\textwidth]{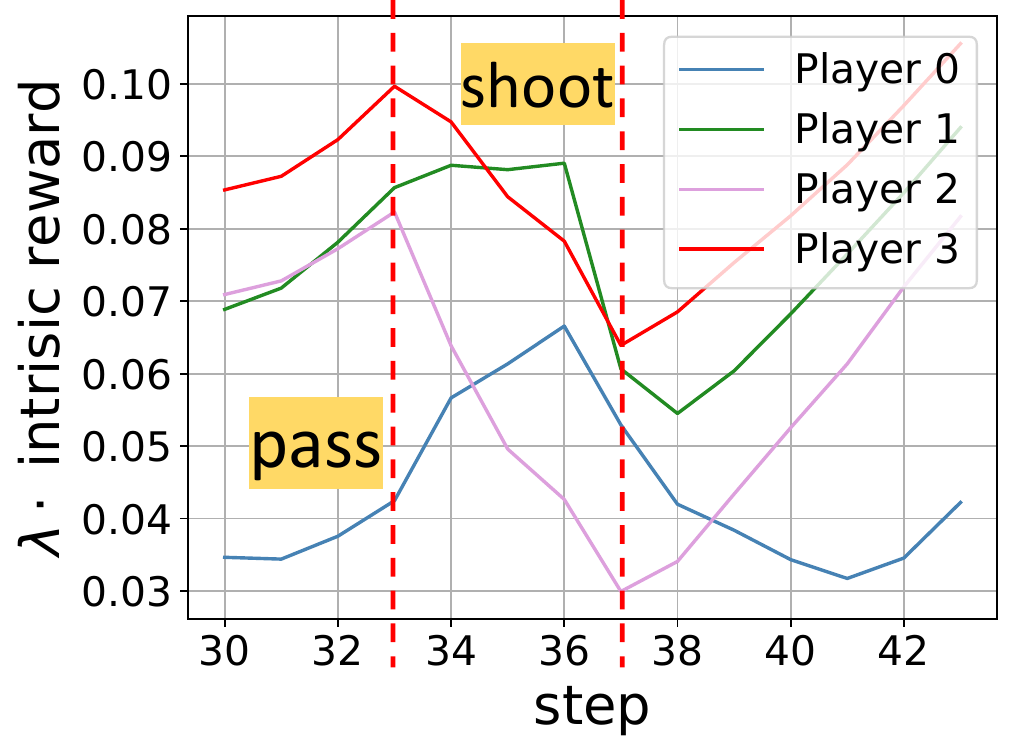}% 图片路径
        % \label{fig:grf_case_learning_curve}
    \end{minipage}
    \caption{\textbf{Left.} Visualization of the trained policy. \textbf{Right.} The learned intrinsic reward curve for the agent players.}
    \label{fig:grf_case}
\end{figure}

The effectiveness of the strategy is demonstrated on the GRF counter\_attack\_hard mission, where we control four players positioned at different locations, cooperating to kick the ball from the half-field area into the opponent's goal. During the course of the play, the team encounters two opposing players controlled by the built-in AI, as well as a goalkeeper defending the opponent's goal. We visualize the trajectories of the players and the ball in Figure~\ref{fig:grf_case} (left), where the players' paths are represented by white arrows and the ball's path by yellow arrows. Agents are distinguished by numerical identifiers. Player 2 dribbles the ball and passes it to player 3, who receives it and shoots it into the goal. Player 1 collaborates with player 2 in the attack, applying pressure on the two opposing defenders, while player 0 moves toward the penalty area to support the offensive play. In Figure~\ref{fig:grf_case} (right), we plot the intrinsic reward curves corresponding to each player during the passing and shooting process.
\begin{itemize}
    \item In the passing process, higher rewards are assigned to player 3 (the player taking the shot), player 2 (the player dribbling the ball), and player 1 (the player supporting the attack). In contrast, player 0, who is farther from the offensive play, receives lower rewards.
    \item After player 2 passes the ball, his intrinsic reward drops sharply. During the flight of the ball, player 1 maintains a relatively high intrinsic reward because he is close to the two opposing defenders and supports the offense. Once player 3 receives the ball and takes the shot, the intrinsic rewards of the three players involved in the offensive play increase significantly.
\end{itemize}
This demonstrates that the intrinsic rewards in our algorithm effectively evaluate each agent's state and its contribution to the overall team performance.

\subsection{Strategy Visualization in State-Reward Space}
\begin{figure}
    \centering
    % 第一张图片
    \begin{minipage}[c]{0.23\textwidth} % 调整宽度
        \centering
        \includegraphics[width=\textwidth]{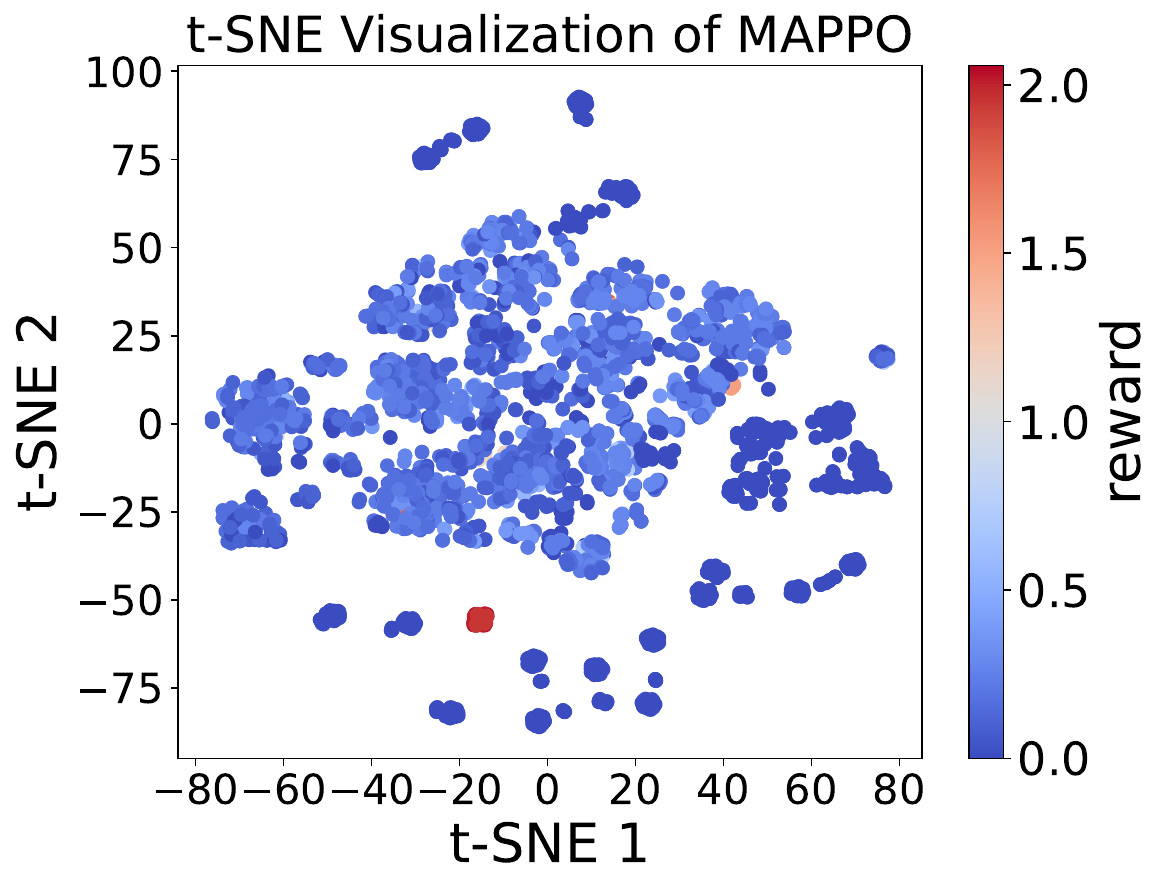} % 图片路径
        % \caption{mappo tsne}
        \label{fig:mappo tsne}
    \end{minipage}
    \hfill % 设置两张图片之间的距离
    % 第二张图片
    \begin{minipage}[c]{0.23\textwidth} % 调整宽度
        \centering
        \includegraphics[width=\textwidth]{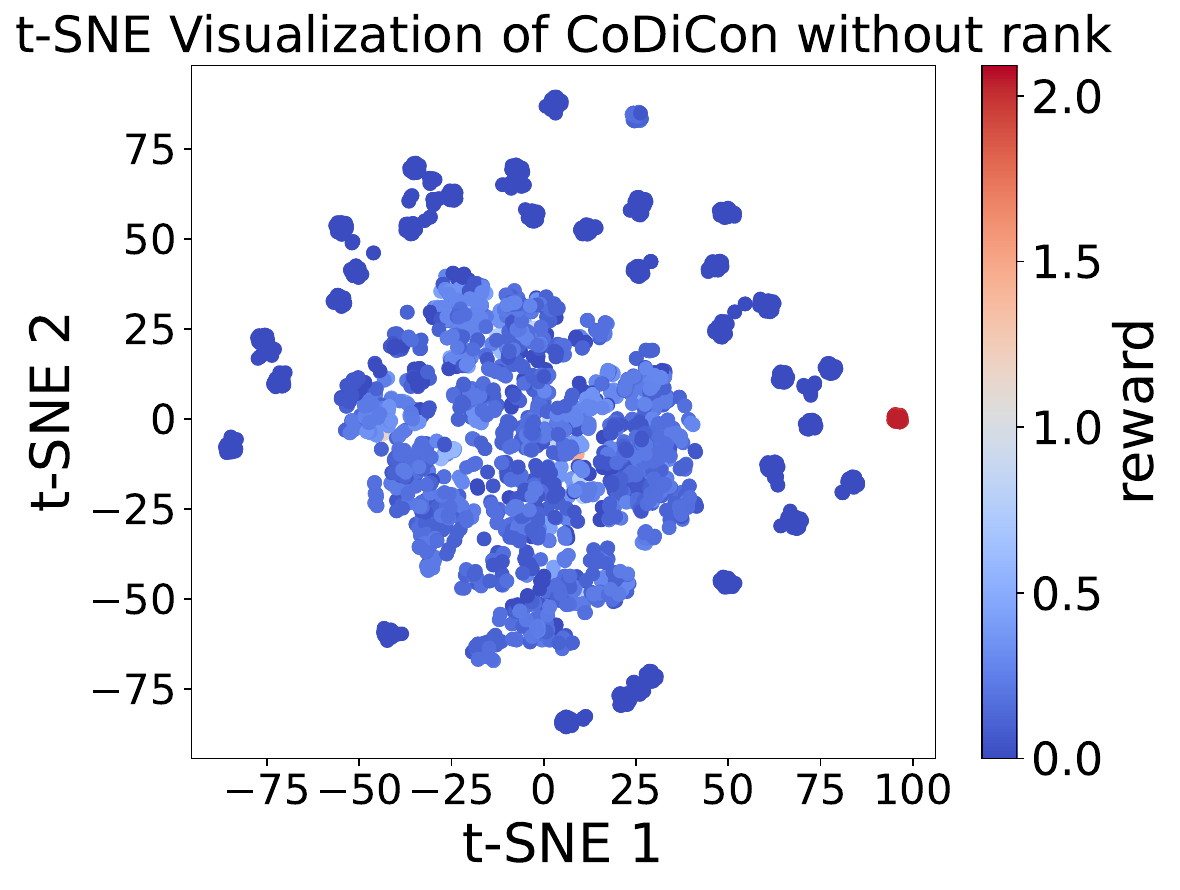} % 图片路径
        % \caption{liir tsne}
        \label{fig:liir tsne}
    \end{minipage}
    \begin{minipage}[c]{0.23\textwidth} % 调整宽度
        \centering
        \includegraphics[width=\textwidth]{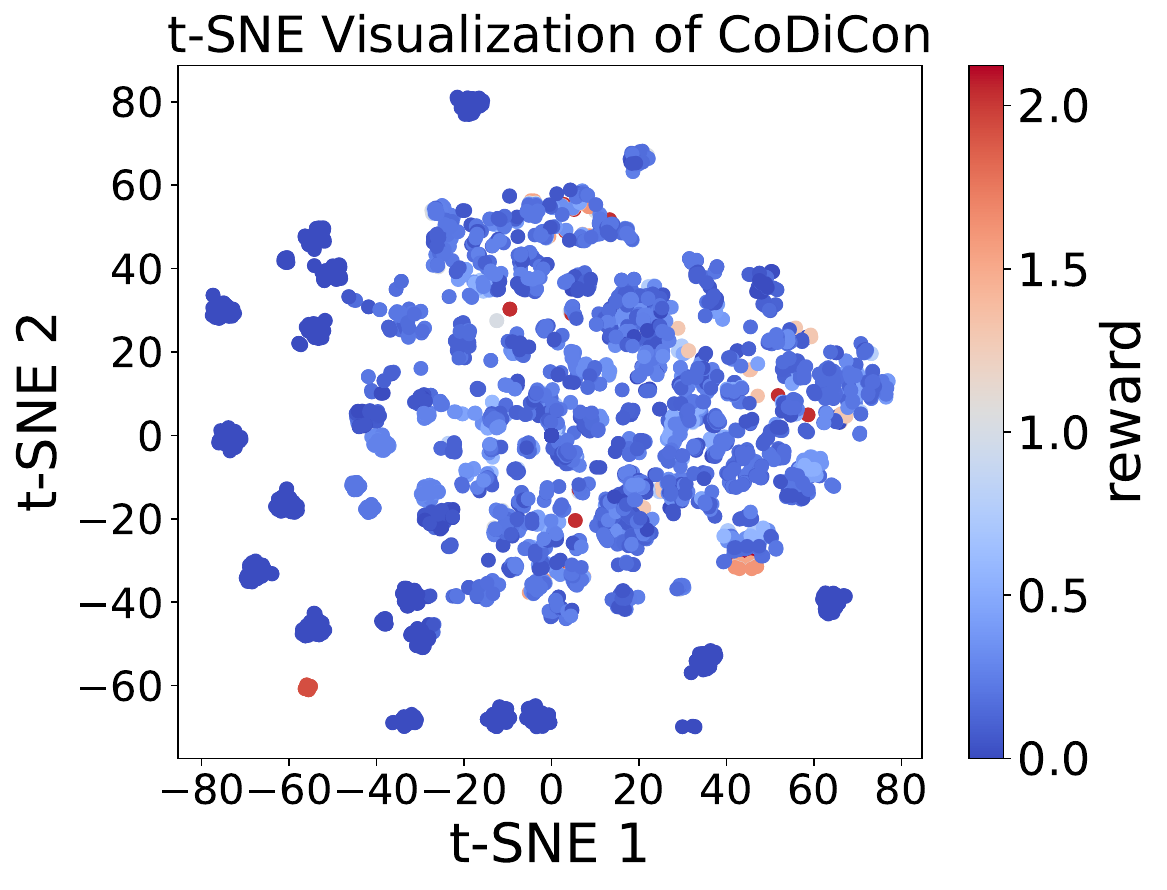} % 图片路径
        % \caption{cc tsne}
        \label{fig:cc tsne}
    \end{minipage}
    \caption{t-SNE results on the state-reward space.}
    \label{fig:tsne}
\end{figure}

In addition to evaluating the performance of the trained policies, we are also interested in understanding how the learned intrinsic reward function influences policy learning. To analyze the intrinsic impact of our algorithm on policy training, we combine the state space and reward space to form a high-dimensional state-reward space. In this space, we can intuitively compare the distribution of the learned policy across states and the rewards obtained by executing actions in these states. We use t-SNE~\cite{van2008visualizing} to perform a low-dimensional mapping of the state-reward space, where the specific value of each point is represented by the environmental reward. Different colors are used on the graph to represent the values, with warmer colors indicating higher values, as shown in the figure~\ref{fig:tsne}. We compared CoDiCon, MAPPO, and the intrinsic reward algorithm without ranking (CoDiCon without rank) based on the results of the trained policies over 50 episodes. As shown in the figure, MAPPO explores more meaningless states and attempts actions over a larger area (larger canvas size) compared to algorithms utilizing intrinsic rewards, but this does not lead to higher rewards. In contrast to the policy without ranked intrinsic rewards, our algorithm avoids meaningless exploration in adjacent feature spaces with high feature similarity (the points are more dispersed yet relatively concentrated) and identifies high-value actions at key nodes (points with more warm colors). The possible reason is that our algorithm facilitates mutual learning among agents for high-reward actions, and the differences in intrinsic rewards make the strategy more complex, diverse, and efficient.

\subsection{Ablation Study}
\begin{figure}
    \centering
    % 第一张图片
    \begin{subfigure}[b]{0.21\textwidth} % 设置宽度
        \centering
        \includegraphics[width=\textwidth]{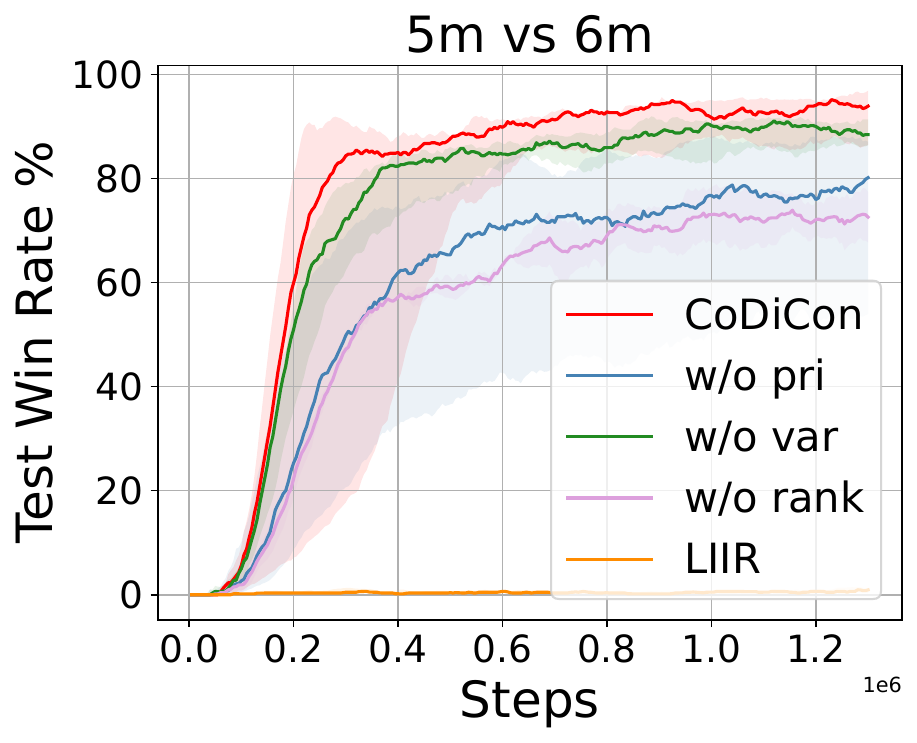} % 图片路径
        % \caption{图片1说明}
    \end{subfigure}
    % 第二张图片
    \begin{subfigure}[b]{0.21\textwidth}
        \centering
        \includegraphics[width=\textwidth]{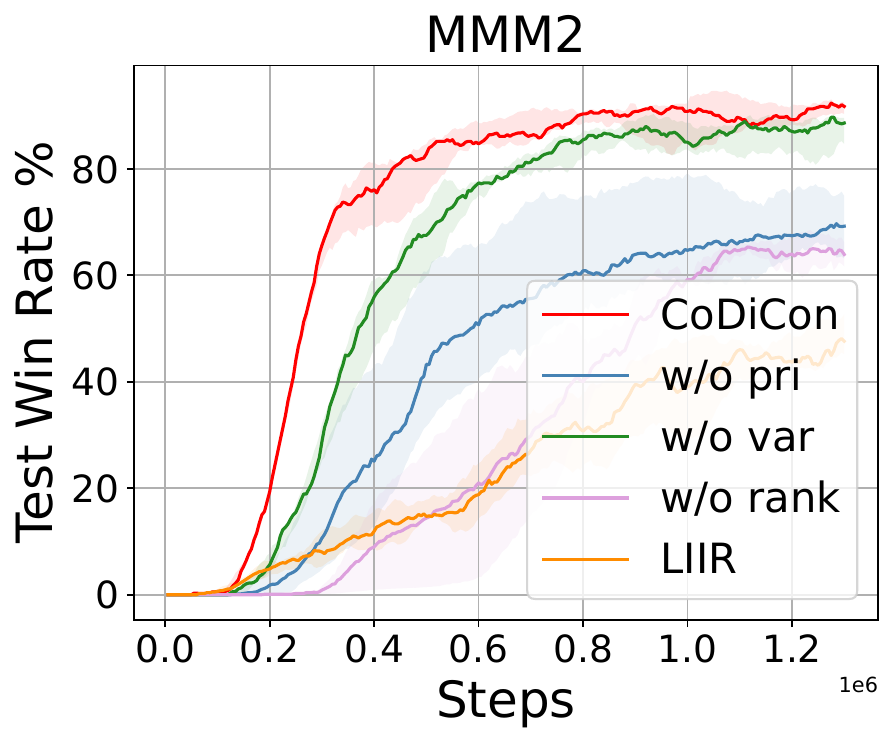} % 图片路径
        % \caption{图片2说明}
    \end{subfigure}    
    \caption{The training curves of success rate for ablation methods.}
    \label{fig:ablation}
\end{figure}

In our algorithm, two loss functions are designed to promote intrinsic reward variability and constrain the training of the ranking module parameters: the priori intrinsic reward MSE loss and the intrinsic reward maximum variance loss. To evaluate the effectiveness of these constraints, we conduct ablation experiments on the two loss functions as shown in Figure~\ref{fig:ablation}. Here, w/o pri refers to the algorithm without the priori loss constraint, w/o var refers to the algorithm without the variance loss constraint, and w/o rank refers to the algorithm without both loss constraints. We perform comparative experiments on the 5m\_vs\_6m and MMM2 scenarios in SMAC, and the results in both scenarios are consistent. These results indicate that both loss functions make positive contributions to policy training, outperforming the performance achieved without the intrinsic reward ranking loss, thereby demonstrating that both constraints are effective. Notably, using only the priori intrinsic reward constraint achieves higher performance and faster convergence compared to using only the variance constraint. A possible explanation is that reinforcement learning involves a large number of low-value states during the early stages of training, where distinguishing these low-value states and assigning them lower intrinsic rewards is critical. In contrast, variance constraints primarily become effective during the later stages of training.
\section{Conclusion}
We propose a novel competitive intrinsic reward multi-agent algorithm, CoDiCon, which enables each agent to learn an intrinsic reward with ranking properties. This approach effectively guides mutual competition and learning among agents, enhancing the efficiency of learning diverse strategies even when the environment provides only a single team reward. Our algorithm is formulated as a constrained bilevel optimization problem, theoretically ensuring that the final optimization objective aligns with maximizing the original environmental rewards. Experimental results on the SMAC, GRF, and Pac-Men environments demonstrate that CoDiCon outperforms existing state-of-the-art methods in terms of performance. Furthermore, case studies further validate the effectiveness of our intrinsic reward design.

\section*{Acknowledgments}

This work is jointly supported by the National Science and Technology Major Project (No.2022ZD0116403), the Key Research Project of Chinese Academy of Sciences (No.RCJJ-145-24-15), the Postdoctoral Fellowship Program of CPSF (Grant No.GZC20232995), the China Postdoctoral Science Foundation (Grant No.2024M763533).

\bibliographystyle{named}
\bibliography{ijcai25}

\end{document}